\documentclass[11pt]{article}

\usepackage[letterpaper, top=1in, bottom=1in, left=1.25in, right=1.25in]{geometry}
\usepackage{graphicx}
\usepackage[hyphens]{url}
\usepackage{amsfonts,amsmath,amstext,amssymb}
\usepackage{enumerate}
\usepackage{tabularx}
\usepackage{mcode}
\usepackage{float}
\usepackage{subcaption}
\usepackage{authblk}

\newtheorem{theorem}{Theorem}
\newtheorem{proposition}{Proposition}
\newtheorem{remark}{Remark}

\newtheorem{opt}{Optimization problem}

\newtheorem{algorithm}{Algorithm}

\newcommand{\bmat}[1]{\begin{bmatrix}#1\end{bmatrix}}
\newcommand\norm[1]{\left\lVert#1\right\rVert}
\newcolumntype{M}[1]{>{\centering\arraybackslash}m{#1}}

\title{\LARGE \bf
Finite Horizon Backward Reachability Analysis and Control Synthesis for Uncertain Nonlinear Systems}
\author[1]{He Yin}
\author[1]{Andrew Packard}
\author[2]{Murat Arcak}
\author[3]{Peter Seiler}
\affil[1]{Department of Mechanical Engineering, University of California, Berkeley}
\affil[2]{Department of Electrical Engineering and Computer Sciences, University of California, Berkeley}
\affil[3]{Department of Aerospace Engineering and Mechanics at the University of Minnesota}

\begin{document}

\maketitle
\thispagestyle{empty}
\pagestyle{empty}

%%%%%%%%%%%%%%%%%%%%%%%%%%%%%%%%%%%%%%%%%%%%%%%%%%%%%%%%%%%%%%%%%%%%%%%%%%%%%%%%
\begin{abstract}
We present a method for synthesizing controllers to steer trajectories from an initial set to a target set on a finite time horizon. The proposed control synthesis problem is decomposed into two steps. The first step under-approximates the backward reachable set (BRS) from the target set, using level sets of storage functions. The storage function is constructed with an iterative algorithm to maximize the volume of the under-approximated BRS. The second step obtains a control law by solving a pointwise min-norm optimization problem using the pre-computed storage function. A closed-form solution of this min-norm optimization can be computed through the KKT conditions. This control synthesis framework is then extended to uncertain nonlinear systems with parametric uncertainties and $\mathcal{L}_2$ disturbances. The computation algorithm for all cases is derived using sum-of-squares (SOS) programming and the S-procedure. The proposed method is applied to several robotics and aircraft examples.
\end{abstract}

%%%%%%%%%%%%%%%%%%%%%%%%%%%%%%%%%%%%%%%%%%%%%%%%%%%%%%%%%%%%%%%%%%%%%%%%%%%%%%%%
\section{INTRODUCTION}
Control synthesis for nonlinear systems suffers from the lack of adequate computational tools. Several recent results leverage sum-of-squares (SOS) and semidefinite programming to construct Lyapunov functions for closed-loop stability and reachability. In \cite{Victor:14} and \cite{Majumdar:14}, a method for designing controllers that maximize backward reachable sets based on occupation measures is proposed, and the synthesis problem is posed as an infinite dimensional linear program, the finite dimensional approximation of which yields a polynomial control policy and an outer approximation of the largest achievable backward reachable set (BRS). However, several Lagrange multipliers are omitted in the optimization formulation, which are of critical importance for achieving tight bounds. Consequently, the outer-approximation of the BRS loses tightness as the state dimension of systems grows, and the control policy is not guaranteed to bring the system to the given target set.

The approach proposed in \cite{Jarvis:05} aims to synthesize control policies to expand the infinite time horizon region of attraction. In \cite{Majumdar:13}, reference tracking controllers are designed to maximize the size of the set of states that are driven to a pre-defined target set. The approach in \cite{Majumdar:17} is to compute a reference tracking controller by minimizing the size of the invariant funnel for the tracking error. An essential advantage of these papers is that, since control laws and storage functions are searched for at the same time, input saturation can be taken into account by adding additional multipliers in the constraints.  On the other hand, since the dependence on decision variables (control policies, storage functions and multipliers) is bilinear, computational algorithms for these methods might be complicated and involve three sub-steps of searching over decision variables.
 
The method presented in \cite{Tan:04} expands the region of attraction certified by a local Control Lyapunov Function (CLF), and control laws are given by variants of the Sontag formula \cite{Sontag:89} \cite{Freeman:95} \cite{Artstein:1983}. The method in \cite{Julian:12} searches for global CLFs whose level sets have similar shapes to those of CLFs obtained from the LQR problem for linearized systems and obtains near optimal performance. The framework is extended to ensure robustness against bounded parameter uncertainties and $\mathcal{L}_2$ disturbances. Other approaches to computational nonlinear control synthesis include: Hamilton-Jacobi methods for reachability computations \cite{Ian:05}, control barrier functions \cite{Aaron:17}, a Lyapunov-based approach utilizing state dependent linear representation of nonlinear systems \cite{Prajna:04}, and a dual to the Lyapunov-based method \cite{Anders:2001}.

This paper addresses the finite time horizon control problem for nonlinear systems that are affine in control, with uncertain parameters and $\mathcal{L}_2$ disturbances. The objective is to maximize the volume of the under-approximated BRSs and to minimize the norm of control inputs that drive trajectories to the target sets. Dissipation inequalities and level sets of storage functions are used to characterize the under-approximated BRSs. The S-procedure \cite{Boyd:94} and SOS for polynomial non-negativity are used to derive the optimization problem of computing storage functions. A computational algorithm is proposed to decompose the optimization problem into convex and quasiconvex subproblems. Since this method does not explicitly search for a control law, the algorithm only involves a two-way search between storage functions and multipliers. Min-norm control laws are given as closed form solutions to quadratic programs based on the computed storage function, and are not restricted to be polynomial functions. 

This paper is a continuation of the stability and reachability analysis methods for nonlinear systems in \cite{He:18} \cite{Ufuk:08} \cite{Weehong:08}, which compute finite and infinite horizon reachable sets and regions of attraction for nonlinear systems with given control laws. In contrast, this paper aims to design controllers as well as under-approximate the BRS.

\section{Notation}
$\mathbb{R}^{m\times n}$ and $\mathbb{S}^n$ denote the set of $m$-by-$n$ real matrices and $n$-by-$n$ real, symmetric matrices. A single superscript index denotes vectors; for example, $\mathbb{R}^m$ is the set of $m \times 1$ vectors whose elements are in $\mathbb{R}$. $\mathcal{C}^1$ is the set of differentiable functions whose derivative is continuous. $\mathcal{L}_2^m$ is the space of $\mathbb{R}^m$-valued measureable functions $f: [0, \infty) \rightarrow \mathbb{R}^m$, with $\norm{f}^2_2 := \int_0^{\infty} f(t)^Tf(t) dt < \infty$. Define $\norm{r}^2_{2,T} := \int_0^T r^T(t)r(t) dt.$ Associated with $\mathcal{L}_2^m$ is the extended space $\mathcal{L}_{2e}^m$, consisting of functions whose truncation $f_T(t) := f(t)$ for $t \le T$; $f_T(t) := 0$ for $t > T$, is in $\mathcal{L}_2^m$ for all $T > 0.$ For $\xi \in \mathbb{R}^n$, $\mathbb{R}[\xi]$ represents the set of polynomials in $\xi$ with real coefficients. The subset $\Sigma[\xi] := \{\pi = \pi_1^2 + \pi_2^2 + ... + \pi_M^2 : \pi_1, ..., \pi_M \in \mathbb{R}[\xi]\}$ of $\mathbb{R}[\xi]$ is the set of SOS polynomials in $\xi$. For $\eta \in \mathbb{R}$, and continuous $r: \mathbb{R}^n \rightarrow \mathbb{R}$, $\Omega_{\eta}^r := \{x \in \mathbb{R}^n : r(x) \le \eta\}.$ For $\eta \in \mathbb{R}$, and continuous $g : \mathbb{R} \times \mathbb{R}^{n} \rightarrow \mathbb{R}$, $\Omega_{t,\eta}^{g} := \{x \in \mathbb{R}^n : g(t,x) \le \eta\}$. 

\section{Storage Function Synthesis}
Consider a single-input, time-varying, nonlinear system with affine dependence on the control input $u$:
\begin{align}
\dot{x} = f(t,x) + g(t,x)u, \label{eq:system1}
\end{align} 
with $x(t) \in \mathbb{R}^n$, $u(t) \in \mathbb{R}$, $f: \mathbb{R}\times \mathbb{R}^n \rightarrow \mathbb{R}^n$, and $g: \mathbb{R}\times \mathbb{R}^n \rightarrow \mathbb{R}^{n}$. Proposition 1 provides conditions on a storage function $V$ and control input $u$ to reach a desired target set.

\begin{proposition}\label{prop1}
	Given system (\ref{eq:system1}), initial time $t_0$, terminal time $T \ge t_0$, and a target set $\Omega_0^{r_T}$, if there exists a $\mathcal{C}^1$ storage function $V: \mathbb{R} \times \mathbb{R}^n \rightarrow \mathbb{R}$ so that
	\begin{align}
	&\inf_{u \in \mathbb{R}} \left\{\frac{\partial V}{\partial t} + \frac{\partial V}{\partial x}\left( f(t,x) + g(t,x)u \right) \right\} \leq 0, \forall (t, x) \in [t_0, T] \times \mathbb{R}^n, \tag{A.1} \label{eq:A1} \\
	&\Omega^V_{T,\gamma} \subseteq \Omega_0^{r_T}, \tag{A.2} \label{eq:A2}
	\end{align}
	then there exists a control law $u = k(t,x)$, such that any trajectory with initial condition $x(t_0) \in \Omega_{t_0, \gamma}^V$ evolves to $x(T) \in \Omega_{0}^{r_T}$, i.e. the final state is in the target set. 
\end{proposition}

The set $\Omega_{t_0, \gamma}^V$ is an under-approximation of the backward reachable set for the given target set and initial time. Proposition \ref{prop1} follows from a simple dissipation argument. Integrating constraint (\ref{eq:A1}) from $t_0$ to $T$ yields $V(T,x(T)) \leq V(t_0,x(t_0))$. Thus it follows from $x(t_0) \in \Omega_{t_0,\gamma}^V$ that $V(T,x(T)) \le \gamma$. Assumption (A.2) then implies that $x(T)$ reaches the target set at time $T$.

   If $\frac{\partial V}{\partial x}g(t,x) \neq 0$ for some $(t,x)$, then (\ref{eq:A1}) is satisfied for $u$ of proper sign and sufficiently large magnitude.  On the other hand, if $\frac{\partial V}{\partial x}g(t,x) = 0$ for some $(t,x)$ then $\frac{\partial V}{\partial t} + \frac{\partial V}{\partial x} f(t,x) \leq 0$ is required to satisfy (\ref{eq:A1}).   Based on this discussion, there exists a control input  $u$ such that (\ref{eq:A1}) is feasible for a given storage function $V$, if and only if the following set containment constraint holds for all $t \in [t_0, T]$:
\begin{align}
 &\left\{x\in \mathbb{R}^n \middle\vert\frac{\partial V}{\partial t} + \frac{\partial V}{\partial x} f(t,x) \leq 0\right\} \supseteq \left\{x\in \mathbb{R}^n \middle\vert \frac{\partial V}{\partial x}g(t,x) = 0 \right\}. \tag{A.3} \label{eq:A3}
\end{align}

\subsection{Local Analysis}
If constraint (\ref{eq:A3}) fails to hold for some points $x$, then we look for a ``local" region that excludes those points. Here we use $\Omega_{t,\gamma}^V$, the $\gamma$ level set of storage function at time $t$, to quantify the local region, and we have the following local version of Proposition \ref{prop1}.
\begin{theorem} \label{thm1}
	Given system (\ref{eq:system1}), initial time $t_0$, terminal time $T \ge t_0$, and a target set $\Omega_0^{r_T}$, if there exists a $\mathcal{C}^1$ storage function $V: \mathbb{R} \times \mathbb{R}^n \rightarrow \mathbb{R}$, such that  $\Omega^V_{T,\gamma} \subseteq \Omega_0^{r_T}$, and for all $t \in [t_0, T]$,  
	\begin{align}
	&\left\{x\in \mathbb{R}^n \middle \vert\frac{\partial V}{\partial t} + \frac{\partial V}{\partial x} f(t,x)\leq 0\right\} \supseteq \left\{x\in \mathbb{R}^n \middle \vert\frac{\partial V}{\partial x}g(t,x) = 0, V(t,x) \leq \gamma \right\}, \tag{B.1} \label{eq:B1} 
	\end{align}
	then there exists a control law $k(t,x)$, such that $x(T) \in \Omega_{0}^{r_T}$, for all $x(t_0) \in \Omega_{t_0, \gamma}^V$. Thus $\Omega_{t_0, \gamma}^V$ is the under-approximation of the backward reachable set for the given target set and initial time. 
\end{theorem}

To find such a storage function using sum-of-squares programming, we restrict $f, g$ and $V$ to polynomial functions. It is often possible to represent nonlinear system equations with polynomials upon changes of variables, Taylor's theorem and least squares regression \cite{Julian:12}. To formulate the set containment constraints, we define the polynomial function $h(t) := (t - t_0)(T - t)$, which is nonnegative when $t \in [t_0, T]$. Since a less conservative under-approximation is preferable, we want to find a storage function $V$ with the volume of $\Omega_{t_0, \gamma}^V$ being maximized. Utilizing the S-procedure to obtain sufficient conditions for the set containment constraints in Theorem \ref{thm1}, and SOS
relaxation for polynomial nonnegativity, we obtain the following optimization problem, with bilinear SOS constraints and a non-convex objective functions.
\begin{opt}\label{opt1}
	\begin{align}
	&\max_{V,l,s} \ \text{volume}(\Omega_{t_0,\gamma}^V) \nonumber \\
	&s.t. \ s_2(t,x), s_3(t,x) \in \Sigma[t,x],  \nonumber \\
	&s_4(x) - \epsilon \in \Sigma[x], \epsilon > 0, l(t,x) \in \mathbb{R}[t,x], \tag{C.1} \label{eq:C1} \\
	&-\left(\frac{\partial V}{\partial t} + \frac{\partial V}{\partial x}f(t,x)\right) - s_2(t,x)h(t) + l(t,x) \frac{\partial V}{\partial x}g(t,x) + s_3(t,x)(V(t,x) - \gamma ) \in \Sigma[t,x], \tag{C.2} \label{eq:C2} \\
	&-s_4(x) r_T(x) + (V(T,x) - \gamma) \in \Sigma[x], \tag{C.3} \label{eq:C3}
	\end{align}
\end{opt}
where the positive number $\epsilon$ ensures that $s_4(x)$ can't take the value of zero.

For bilinear SOS constraints (\ref{eq:C1}) to (\ref{eq:C3}), $l(t,x)$ and $\frac{\partial V}{\partial x}$, $s_3(t,x)$ and $V(t,x)$ are two pairs of bilinear decision variables. To tackle this non-convex optimization problem, we decompose it into two subproblems to iteratively search between storage function $V$ and multipliers $s,l$. 
\begin{algorithm} \label{alg1}
	Iterative method \newline
	\textbf{Inputs:} A storage function $V^0$ satisfying constraints (\ref{eq:C2}) and (\ref{eq:C3}).\newline
	\textbf{Outputs:} $\Omega_{t_0,\gamma}^V$ with its volume maximized.
	\begin{enumerate}
		\item $\gamma$ step: maximization problem
		\begin{align}
		&\max_{\gamma, l, s_2, s_3, s_4} \gamma \nonumber \\
		&s.t. \ s_2(t,x), s_3(t,x) \in \Sigma[t,x], l(t,x) \in \mathbb{R}[t,x]  \nonumber \\
		& s_4(x)-\epsilon \in \Sigma[t,x], \epsilon > 0, \nonumber \\
		&-\left(\frac{\partial V^0}{\partial t} + \frac{\partial V^0}{\partial x}f(t,x)\right) - s_2(t,x)h(t) + l(t,x) \frac{\partial V^0}{\partial x}g(t,x) + s_3(t,x)(V^0(t,x) - \gamma ) \in \Sigma[t,x], \nonumber \\
		&-s_4(x) r_T(x) + (V^0(T,x) - \gamma) \in \Sigma[x]. \nonumber 
		\end{align}
		\item $V$ step: feasibility problem over decision variables $V, s_1, s_2, s_4$
		\begin{align}
		& s_1(x) \in \Sigma[x], s_2(t,x) \in \Sigma[t,x], \nonumber \\
		& s_4(x)-\epsilon \in \Sigma[t,x], \epsilon > 0, \nonumber \\
		& -(V(t_0,x) - \gamma^*) + s_1(x) (V^0(t_0,x) - \gamma^*), \nonumber \\
		&-\left(\frac{\partial V}{\partial t} + \frac{\partial V}{\partial x}f(t,x)\right) - s_2(t,x)h(t) +\bar{l}(t,x) \frac{\partial V}{\partial x}g(t,x) + \bar{s}_3(t,x)(V(t,x) - \gamma^* ) \in \Sigma[t,x], \nonumber \\
		&-s_4(x) r_T(x) + (V(T,x) - \gamma^*) \in \Sigma[x]. \nonumber 
		\end{align}
	\end{enumerate}
\end{algorithm}
\begin{remark}
	For the $\gamma$ step, $V^0$ is the storage function computed from the $V$ step of the previous iteration. Since $s_3$ and $\gamma$ enter bilinearly, and $\gamma$ is the objective function, then the $\gamma$ step is a generalized SOS problem, which is proven in \cite{Pete:10} to be quasiconvex. Thus, the global optimal solution can be computed by bisecting $\gamma$.  
\end{remark}

\begin{remark}
	$\bar{l}$, $\bar{s}_3$ and $\gamma^*$ in the $V$ step are obtained from the $\gamma$ step. Similar to the algorithm proposed in \cite{Iannelli:18} to find the region of attraction, this algorithm makes use of $V^0$ from the previous iteration as a shape function for enlarging the volume of $\Omega_{t_0,\gamma}^V$, rather than using a preset shape function. After the $\gamma$ step, constraints of the $\gamma$ step are active for $V^0$. In the $V$ step, a new feasible $V$ is computed, which is the analytic center of the LMI constraints. Thus the $V$ step feasibility problem pushes $V$ away from the constraints, which give the next $\gamma$ step more freedom to increase $\gamma$. The $V$ step is a SOS problem, which is convex. Note that although global optima for the subproblems in the $\gamma$ and $V$ steps at each iteration can be achieved, the ultimate solution of this iterative algorithm is not necessarily the global optimal solution for optimization problem \ref{opt1}.
\end{remark}

\begin{remark}
	Since in many cases, we want to bring the system close to an equilibrium point, the target region is set as a neighborhood around it. Therefore, LQR controllers designed for linearization of dynamics about equilibrium points can be used to compute storage functions, which can be used to initialize $V^0$.
\end{remark}

\subsection{Multi-input Case}
In this section, the framework in the previous sections is extended to multi-input systems. Assume that there are $m$ inputs $u \in \mathbb{R}^m$, and accordingly $g: \mathbb{R}\times \mathbb{R}^n \rightarrow \mathbb{R}^{n \times m}$. Denote $g = \left[g_1, g_2, ... , g_m\right]$, where $g_i$ is the $i^{th}$ column of $g$; denote $u = [u_1, u_2, ... , u_m]^T$ and write the multi-input system as 
\begin{align}
\dot{x} = f(t,x) + \sum_{i=1}^m g_i(t,x) u_i. \label{eq:system2}
\end{align}

The constraint (\ref{eq:B1}) is modified to be, for all $t \in [t_0, T]$,
\begin{align}
&\left\{ x \in \mathbb{R}^n \middle\vert \frac{\partial V}{\partial t} + \frac{\partial V}{\partial x}f(t,x)\leq 0\right\} \supseteq \left\{ x \in \mathbb{R}^n \middle\vert \frac{\partial V}{\partial x}g_1(t,x) = 0, ... , \frac{\partial V}{\partial x}g_m(t,x) = 0, V(t,x) \leq \gamma \right\}.\tag{D.1} \label{eq:D1}
\end{align}
Applying the S-procedure to (\ref{eq:D1}), we have its corresponding SOS constraint
\begin{align}
&-\left(\frac{\partial V}{\partial t} + \frac{\partial V}{\partial x}f(t,x)\right) - s_2(t,x)h(t) + \sum_{i=1}^m \bigg\{l_i(t,x) \frac{\partial V}{\partial x}g_i(t,x)\bigg\} + s_3(t,x)(V(t,x) - \gamma ) \in \Sigma[t,x]. \tag{D.2} \label{eq:D2}
\end{align}
By replacing (\ref{eq:C2}) with (\ref{eq:D2}), and keeping other constraints to be the same, we obtain an optimization problem for multi-input systems. Instead of only searching over $l$, we now search over polynomials $l_i, i = 1, ..., m$.

\section{Min-norm Control Synthesis}
With the storage function $V$ computed from optimization problem \ref{opt1}, we want to find a control law $k: [t_0, T]\times \mathbb{R}^n \rightarrow \mathbb{R}^m$, such that the dissipation inequality in constraint (\ref{eq:A1}) holds for all $x \in \Omega_{t,\gamma}^V$ and $t \in [t_0, T]$. Also, to avoid excessive control magnitudes, we want the norm of $u$ to be minimized. Similar to the idea in \cite{Freeman:95}, the control input $u$ is determined by solving the following quadratic program (QP),
\begin{equation}
\begin{aligned}
&\min_{u \in \mathbb{R}^m} u^T u  \\
&s.t. \ \ \ \frac{\partial V(t,x)}{\partial t} + \frac{\partial V(t,x)}{\partial x} \left(f(t,x) + g(t,x)u\right) \leq 0. \label{eq:MinNormOpt}
\end{aligned}
\end{equation}

Since the QP (\ref{eq:MinNormOpt}) satisfies Slater's condition, its closed form solution can be obtained by solving the KKT condition, which yields the optimal  control law
\begin{align}
k(t,x) = u^* =
\begin{cases}
0,& b(t,x) \le 0 \\
\frac{-b(t,x)}{a(t,x) a(t,x)^T}a(t,x)^T, & b(t,x) > 0, \label{eq:controlLaw}
\end{cases}
\end{align} 

where 
\begin{equation}
\begin{aligned}
&a(t,x) := \frac{\partial V(t,x)}{\partial x} g(t,x), \label{eq:ab1} \\
&b(t,x) := \frac{\partial V(t,x)}{\partial t} + \frac{\partial V(t,x)}{\partial x} f(t,x). 
\end{aligned}
\end{equation}

\begin{remark}
	The constraint (\ref{eq:B1}): $a(t,x) = 0$ implies $b(t,x) \leq 0$, ensures that if $b(t,x)>0$, we have $a(t,x) \neq 0$. Therefore, there is no singularity in the control law due to division by $a(t,x)a(t,x)^T$. However, discontinuity in the control law might be possible at the points $(t,x)$, where $b(t,x) = 0$ and $a(t,x) = 0$. To deal with discontinuity, we can use a strict version of constraint (\ref{eq:B1}): $a(t,x) = 0$ implies $b(t,x) < 0$, for all $x \in \Omega_{t,\gamma}^V\backslash \bar{x}$, for all $t \in [t_0, T]$ and $a(t,\bar{x})=0$ implies $b(t,\bar{x}) \leq 0$ for all $t \in [t_0, T]$ , where $\bar{x}$ can be the origin or some equilibrium point for the system. Then discontinuity can only happen at the point $\bar{x}$, and continuity of the control law at $\bar{x}$ can be established using an analog of the small control property from \cite{Sontag:89}.
\end{remark}

\section{Modifications for Systems with Bounded Uncertainties}
For brevity of notation, we still consider single-input systems, but with uncertain parameters $\delta$,
\begin{align}
\dot{x} = F(t,x,\delta) + g(t,x)u, \label{eq:system3}
\end{align}
with $x(t) \in \mathbb{R}^n$, $u(t) \in \mathbb{R}$, $\delta(t) \in \mathbb{R}^{n_{\delta}}$, $F: \mathbb{R}\times \mathbb{R}^n \times \mathbb{R}^{n_{\delta}} \rightarrow \mathbb{R}^n$,  $g: \mathbb{R}\times \mathbb{R}^n \rightarrow \mathbb{R}^{n}$, and assume that $\delta$ lies in a known set $\Delta$
\begin{align}
\delta \in \Delta := \{\delta \in \mathbb{R}^{n_\delta}| N(\delta) \ge 0\}. \nonumber
\end{align}

Slightly modifying constraint ($\ref{eq:B1}$), we have the dissipation inequality constraint for the uncertain system: for all $(t,\delta) \in [t_0,T]\times\Delta$,
\begin{align}
&\left\{ x \in \mathbb{R}^n \middle\vert \frac{\partial V}{\partial t} + \frac{\partial V}{\partial x} F(t,x,\delta) \leq 0\right\} \supseteq \left\{x \in \mathbb{R}^n \middle\vert \frac{\partial V}{\partial x}g(t,x) = 0, V(t,x) \leq \gamma\right\}. \tag{E.1} \label{eq:E1}
\end{align}

Assume that $F(t,x,\delta)$ is affine in $\delta$, and denote $F(t,x,\delta) = f(t,x) + g_{\delta}(t,x)\delta$, with $g_{\delta} :\mathbb{R}\times \mathbb{R}^n \rightarrow \mathbb{R}^{n \times n_{\delta}} $.
To simplify the analysis, assume also that the set $\Delta$ is a bounded polytope, and define the set of vertices of $\Delta$, $\mathcal{E}_{\Delta}:= \{\delta^{[1]}, \delta^{[2]},..., \delta^{[N_{vertex}]}\}$, where $N_{vertex}$ is the number of vertices. Since $\delta$ enters the system linearly, and it lies in a bounded polytope, if we impose constraint (\ref{eq:E1}) to hold on $\mathcal{E}_{\Delta}$, then it holds everywhere on $\Delta$. Then constraint (\ref{eq:E1}) can be transformed into a number of $N_{vertex}$ constraints, for all $t \in [t_0, T]$,
\begin{align}
&\left\{x\in \mathbb{R}^n \middle\vert \frac{\partial V}{\partial t} + \frac{\partial V}{\partial x} f(t,x) + \frac{\partial V}{\partial x} g_{\delta}(t,x)\delta^{[i]} \leq 0 \right\} \supseteq \nonumber \\
&\quad \quad \quad \left\{x\in \mathbb{R}^n \middle\vert \frac{\partial V}{\partial x}g(t,x) = 0, V(t,x) \leq \gamma \right\}, \forall i = 1, ... , N_{vertex}. \tag{E.2} \label{eq:E2}
\end{align}
Note that constraint (\ref{eq:E2}) doesn't introduce $\delta$ as a new variable, which helps to reduce computation time.

\subsection{Control Synthesis for Systems with Bounded Uncertainties}
Similar to the QP (\ref{eq:MinNormOpt}), we have the min-norm QP for the uncertain system
\begin{equation}
\begin{aligned}
&\min_{u \in \mathbb{R}^m} u^T u  \\
&s.t. \ \frac{\partial V(t,x)}{\partial t} + \frac{\partial V(t,x)}{\partial x} \big(f(t,x) + g(t,x)u + g_{\delta}(t,x)\delta^{[i]}\big) \leq 0, \ \forall i = 1,..., N_{vertex}. \label{eq:MinNormOpt2}
\end{aligned}
\end{equation}

Define 
\begin{equation}
\begin{aligned}
&a(t,x) := \frac{\partial V}{\partial x}g(t,x), \label{eq:ab2} \\
&b_i(t,x) := \frac{\partial V}{\partial t} + \frac{\partial V}{\partial x}f(t,x)+ \frac{\partial V}{\partial x}g_{\delta}(t,x)\delta^{[i]}, \\
&b_{max}(t,x) := \max\{b_1(t,x),...,b_{N_{vertex}}(t,x)\}. 
\end{aligned}
\end{equation}

Then the QP (\ref{eq:MinNormOpt2}) can be rewritten as 
\begin{equation}
\begin{aligned}
&\min_{u \in \mathbb{R}^m} u^T u  \\
&s.t. \ a(t,x)u + b_{i}(t,x) \leq 0, \ \forall i =1,...,N_{vertex}, \nonumber
\end{aligned}
\end{equation}
which is equivalent to
\begin{equation}
\begin{aligned}
&\min_{u \in \mathbb{R}^m} u^T u  \\
&s.t. \ a(t,x)u + b_{max}(t,x) \leq 0. \label{eq:MinNormOpt3}
\end{aligned}
\end{equation}
The control law is given by 
\begin{align}
k(t,x) = u^* =
\begin{cases}
0,& b_{max}(t,x) \le 0 \\
\frac{-b_{max}(t,x)}{a(t,x) a(t,x)^T}a(t,x)^T, & b_{max}(t,x) > 0 \label{eq:controlLaw2}
\end{cases}.
\end{align}

\section{Modifications for Systems with $\mathcal{L}_2$ Disturbances}
Consider a disturbed system with disturbances $w$ entering linearly
\begin{align}
\dot{x} = f(t,x) + g(t,x)u + g_{w}(t,x)w, \label{eq:system4}
\end{align}
with $x(t) \in \mathbb{R}^n$, $u(t) \in \mathbb{R}$, $w(t) \in \mathbb{R}^{n_{w}}$, $f: \mathbb{R}\times \mathbb{R}^n \rightarrow \mathbb{R}^n$,  $g: \mathbb{R}\times \mathbb{R}^n \rightarrow \mathbb{R}^{n}$, and $g_w :\mathbb{R}\times \mathbb{R}^n \rightarrow \mathbb{R}^{n \times n_{w}} $.

\begin{theorem} \label{thm2}
	Given system (\ref{eq:system4}), initial time $t_0$, terminal time $T \ge t_0$, a target set $\Omega_0^{r_T}$, and disturbances $w$ satisfying $\int_{t_0}^t w(\tau)^T w(\tau) d\tau \le R^2 q(t)$, where the non-decreasing polynomial function $q$ satisfies $q(t_0)=0$, $q(T)=1$, if there exists a $\mathcal{C}^1$ storage function $V: \mathbb{R} \times \mathbb{R}^n \rightarrow \mathbb{R}$, satisfying $\Omega^V_{T,\gamma+R^2} \subseteq \Omega_0^{r_T}$, and for all $(t,w) \in [t_0, T]\times\mathbb{R}^w$, such that 
	\begin{align}
	&\left\{x \in \mathbb{R}^n \middle\vert \frac{\partial V}{\partial t} + \frac{\partial V}{\partial x} f(t,x) + \frac{\partial V}{\partial x}g_w(t,x) w \leq w^T w \right\} \supseteq  \nonumber \\
	&\quad \quad \quad \left\{x \in \mathbb{R}^n \middle\vert \frac{\partial V}{\partial x}g(t,x) = 0, V(t,x) \leq \gamma + R^2 q(t) \right\}, \tag{F.1} \label{eq:F1} 
	\end{align}
	then there exists a control law $k(t,x)$, such that $x(T) \in \Omega_{0}^{r_T}$, for all $x(t_0) \in \Omega_{t_0, \gamma}^V$. 
\end{theorem}

In Theorem \ref{thm2}, the function $q$ describes how fast the energy of disturbances releases. If $q$ is not known beforehand, we need to relax constraint (\ref{eq:F1}) to be: for all $(t,w) \in [t_0, T]\times\mathbb{R}^w$,
\begin{align}
&\left\{x \in \mathbb{R}^n \middle\vert \frac{\partial V}{\partial t} + \frac{\partial V}{\partial x} f(t,x) + \frac{\partial V}{\partial x}g_w(t,x) w \leq w^T w \right\} \supseteq  \nonumber \\
&\quad \quad \quad \quad \left\{x \in \mathbb{R}^n \middle\vert \frac{\partial V}{\partial x}g(t,x) = 0, V(t,x) \leq \gamma + R^2 \right\}, \tag{G.1} \label{eq:G1} 
\end{align}
which can be more restrictive for the storage function, since the dissipation inequality is required to hold on a larger space in $x$.

If, in addition to the $\mathcal{L}_2$ bound above, we have a $\mathcal{L}_{\infty}$ constraint for $w: w(t)^Tw(t) \leq \alpha$, for all $t \in [t_0, T]$. Constraint (\ref{eq:F1}) in Theorem \ref{thm2} is modified to hold for all $(t, w) \in [t_0, T] \times \left\{w \in \mathbb{R}^w \middle\vert w^Tw \leq \alpha \right\}$. By modifying the SOS constraint (\ref{eq:C2}), the SOS constraint for (\ref{eq:F1}) can be written as 
\begin{align}
&-\left(\frac{\partial V}{\partial t} + \frac{\partial V}{\partial x}f(t,x) + \frac{\partial V}{\partial x}g_w(t,x)w - w^Tw\right)  + l(t,x,w) \frac{\partial V}{\partial x}g(t,x) - s_a(t,x,w)h(t) + \nonumber \\
 &\quad \quad \quad \quad s_b(t,x,w)(V(t,x) - \gamma - R^2q(t)) + s_c(t,x,w)(w^Tw - \alpha) \in \Sigma[t,x,w],
\end{align}
where $s_a(t,x,w), s_b(t,x,w), s_c(t,x,w) \in \Sigma[t,x,w]$, $l(t,x,w) \in \mathbb{R}[t,x,w]$.

\subsection{Control Synthesis for Disturbed Systems}
Similar to the QP (\ref{eq:MinNormOpt}), the following QP gives a min-norm control input for the disturbed system, assuming the value of $w$ is not accessible
\begin{equation}
\begin{aligned}
&\min_{u \in \mathbb{R}^m} u^T u  \\
&s.t. \ \frac{\partial V(t,x)}{\partial t} + \frac{\partial V(t,x)}{\partial x} (f(t,x) + g(t,x)u +  g_w(t,x)w ) \leq w^T w, \forall w \in \{w \in \mathbb{R}^w | w^T w \leq \alpha\}. \label{eq:MinNormOpt4}
\end{aligned}
\end{equation}

For brevity of notation, define $c(t,x) := \left(\frac{\partial V(t,x)}{\partial x}g_w(t,x)\right)^T$, $d(t,x) := \frac{\partial V(t,x)}{\partial t} + \frac{\partial V(t,x)}{\partial x}f(t,x) + \frac{\partial V(t,x)}{\partial x}g(t,x)u$. The constraint in QP (\ref{eq:MinNormOpt4}) can then be restated as 
\begin{align}
&\max_{w \in \{w \in \mathbb{R}^w | w^T w \leq \alpha\}}\left(- w^T w + c^Tw + d\right)\leq 0. \label{eq:wconstr1}
\end{align}
Solving it with the KKT condition, we have 
\begin{align}
w^* = 
\begin{cases}
\frac{\sqrt{\alpha}}{\sqrt{c^T c}}c, & c^T c \ge 4 \alpha,  \\
\frac{1}{2}c, & c^T c < 4 \alpha. \nonumber
\end{cases}
\end{align} 

Substituting $w^*$ into optimization problem (\ref{eq:MinNormOpt4}), we get two QPs for two cases. The formula of control law for disturbed systems is the solution to QPs, and it is the same as equation (\ref{eq:controlLaw}), whereas $a(t,x)$ and $b(t,x)$ are 
\begin{equation}
\begin{aligned}
&a(t,x) := \frac{\partial V(t,x)}{\partial x} g(t,x),  \\
&b(t,x) :=  \\
&\begin{cases}  
\frac{\partial V}{\partial t} + \frac{\partial V}{\partial x} f(t,x) + \sqrt{\alpha c^T c} - \alpha, & c^T c \ge 4 \alpha, \\
\frac{\partial V}{\partial t} + \frac{\partial V}{\partial x} f(t,x) + \frac{c^T c}{4}, & c^T c<4 \alpha. \label{eq:ab3} 
\end{cases}
\end{aligned}
\end{equation}

\section{Control Synthesis for Disturbed Systems with Bounded Uncertainties}
Consider a system with both parametric uncertainties $\delta$ and disturbances $w$
\begin{align}
&\dot{x}(t) = f(t,x) + g(t,x)u + g_{w}(t,x)w +  g_{\delta}(t,x)\delta. \label{eq:system5}
\end{align}
Again, we assume that $\delta$ lies in the bounded polytope $\Delta$, and slightly modifying constraint (\ref{eq:F1}), we get the dissipation inequality for system (\ref{eq:system5}), for all $(t, w) \in [t_0, T] \times \{w \in \mathbb{R}^w | w^Tw \leq \alpha\}$,
\begin{align}
&\bigg\{ x \in \mathbb{R}^n \bigg\vert \frac{\partial V}{\partial t} + \frac{\partial V}{\partial x} f(t,x) + \frac{\partial V}{\partial x}g_w(t,x) w + \frac{\partial V}{\partial x}g_{\delta}(t,x)\delta^{[i]} \leq w^T w\bigg\} \supseteq \bigg\{x \in \mathbb{R}^n \bigg\vert \frac{\partial V}{\partial x}g(t,x) = 0,  \nonumber \\
&V(t,x) \leq \gamma + R^2 q(t) \bigg\}, \forall i = 1,..., N_{vertex}. \tag{H.1} \label{eq:H1}
\end{align}

After a storage function $V$ is obtained, the control input is computed through the following QP
\begin{equation}
\begin{aligned}
&\min_{u \in \mathbb{R}^m} u^T u  \\
&s.t. \ \frac{\partial V(t,x)}{\partial t} + \frac{\partial V(t,x)}{\partial x} \big(f(t,x) + g(t,x)u +  g_w(t,x)w\\
 &\quad \quad + g_{\delta}(t,x)\delta^{[i]}\big) \leq w^T w, \forall w \in \{w \in \mathbb{R}^w | w^T w \leq \alpha\}, \forall i = 1, ..., N_{vertex}. \label{eq:MinNormOpt5}
\end{aligned}
\end{equation}

Define $e_i(t,x) := \frac{\partial V(t,x)}{\partial x}g_{\delta}(t,x)\delta^{[i]}$ and $e_{max} := \max\{e_1, ..., e_{N_{vertex}}\}$. The constraint in QP (\ref{eq:MinNormOpt5}) can be restated as 
\begin{align}
&\max_{w \in \{w \in \mathbb{R}^w | w^T w \leq \alpha\}}\left(- w^T w + c^Tw + d + e_i\right)\leq 0, \forall i = 1, ..., N_{vertex}, \nonumber 
\end{align}
which is equivalent to 
\begin{align}
&\max_{w \in \{w \in \mathbb{R}^w | w^T w \leq \alpha\}}\left(- w^T w + c^Tw + d + e_{max}\right)\leq 0. \label{eq:wconstr2}  
\end{align}
Notice that constraints (\ref{eq:wconstr1}) and (\ref{eq:wconstr2}) has the same optimal solution $w^*$. Substituting $w^*$ back into constraint (\ref{eq:wconstr2}), we have two QPs. The formula of control law is the solution to QPs, and it is the same as equation (\ref{eq:controlLaw}), whereas $a(t,x)$ and $b(t,x)$ are
\begin{equation}
\begin{aligned}
&a(t,x) := \frac{\partial V(t,x)}{\partial x} g(t,x),  \\
&b(t,x) :=  \\
&\begin{cases}  
\frac{\partial V}{\partial t} + \frac{\partial V}{\partial x} f(t,x) + \sqrt{\alpha c^T c} - \alpha + e_{max}, & c^T c \ge 4 \alpha, \\
\frac{\partial V}{\partial t} + \frac{\partial V}{\partial x} f(t,x) + \frac{c^T c}{4} + e_{max}, & c^T c<4 \alpha.  \label{eq:ab4}
\end{cases}
\end{aligned}
\end{equation}

\section{Examples}
A workstation with four 2.7 [GHz] Intel Core i5 64 bit processors and 8[GB] of RAM was
used for performing all computations in the following examples. The SOS optimization
problem is formulated and translated into SDP using the sum-of-square module in SOSOPT \cite{Pete:13} on MATLAB, and solved by the SDP solver Mosek \cite{Mosek:17}. Table \ref{tab:table} shows the degree of
polynomials we chose, and the computation time it took for each example.
\begin{table}[h]
	\caption{Computation times for each example \label{tab:table}}
	\centering
	\begin{tabular}{ |M{2.8cm}|M{1.8cm}|M{1.8cm}|M{1.8cm}|M{1.8cm}|M{1.8cm}|}
		\hline
		Examples & Number of States &Degree of Dynamics &Degree of $V(t,x)$ &Degree of $s, l$ &Computing Time [sec] \\
		\hline
		Section \ref{ex1}: 2-state example    &2 & 3 &6 & 6 & $7.2 \times 10^2$\\ \hline 
		Section \ref{ex2}: Dubin's car    &3 &1 & 4 &4  & $2.5 \times 10^3$   \\ \hline
		Section \ref{ex3}: Cart-pole      &4 & 5 &4 &4 &$5.3 \times 10^3$\\ \hline
		Section \ref{ex4}: Pendubot   &4&3&4&4&$1.5 \times 10^3$ \\ \hline
		Section \ref{ex5}: GTM   &4&3&4&4& $1.8 \times 10^3$ \\ \hline
		\end{tabular}
\end{table}
\subsection{Uncertain Two-State Example} \label{ex1}
Consider the following uncertain two-state dynamics from \cite{Jarvis:05}, where a parametric uncertainty $\delta$ enters the system linearly
\begin{equation}
\dot{x}_1 = u, \ \dot{x}_2 = -x_1 + \frac{1}{6}x_1^3 \delta - u, \label{eq:system6}
\end{equation}
with the prior knowledge that $\delta \in \Delta := [-1.1, 1.2]$.

\begin{figure}[h]
	\centering
	\includegraphics[width=0.4\textwidth]{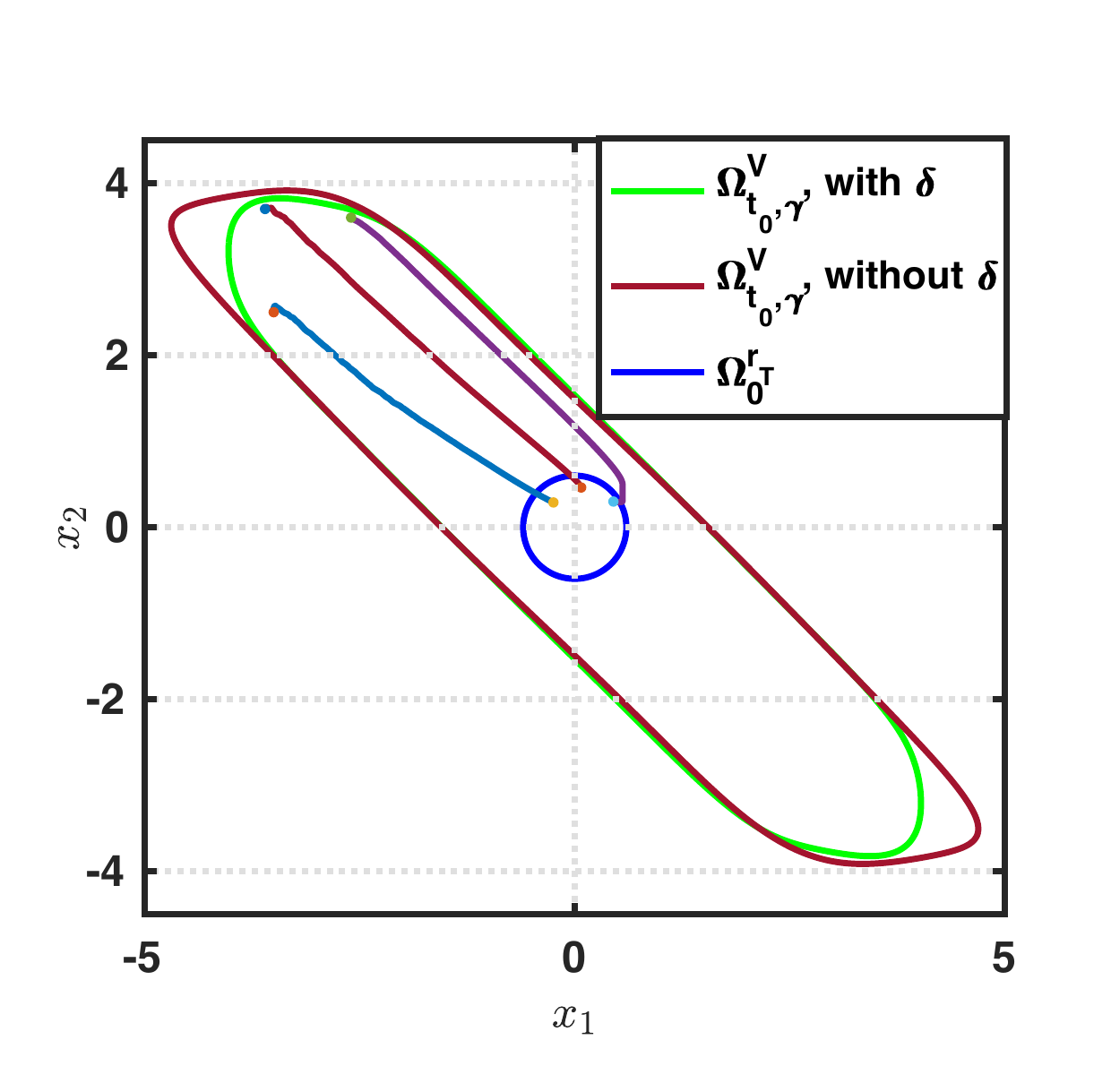}
	\caption{Uncertain two-state example}
	\label{fig:2stateUncertain}    
\end{figure}

The time horizon is chosen to be $[t_0, T] = [0, 1 \ \text{sec}]$, and the target set is given to be $\Omega_{0}^{r_T} = \{x \in \mathbb{R}^2 | x^T$ $diag(1/0.6^2, 1/0.6^2) x - 1 \le 0\}$, which is shown as the blue circle in Figure \ref{fig:2stateUncertain}. $\Omega_{t_0,\gamma}^V$ for the uncertain system is shown with the green curve in Figure \ref{fig:2stateUncertain}, and the brown curve is $\Omega_{t_0,\gamma}^V$ for the system with $\delta$ set to be $1$, i.e. for the system without uncertainty. The three trajectories are simulations of system (\ref{eq:system6}), with the control law defined by equations (\ref{eq:ab2})(\ref{eq:controlLaw2}) and uncertain parameter $\delta(t)$ drawn from the uniform distribution on $[-1.1, 1.2]$ at each time step,

\subsection{Dubin's Car} \label{ex2}
\begin{figure}[h]
	\centering
	\includegraphics[width=0.7\textwidth]{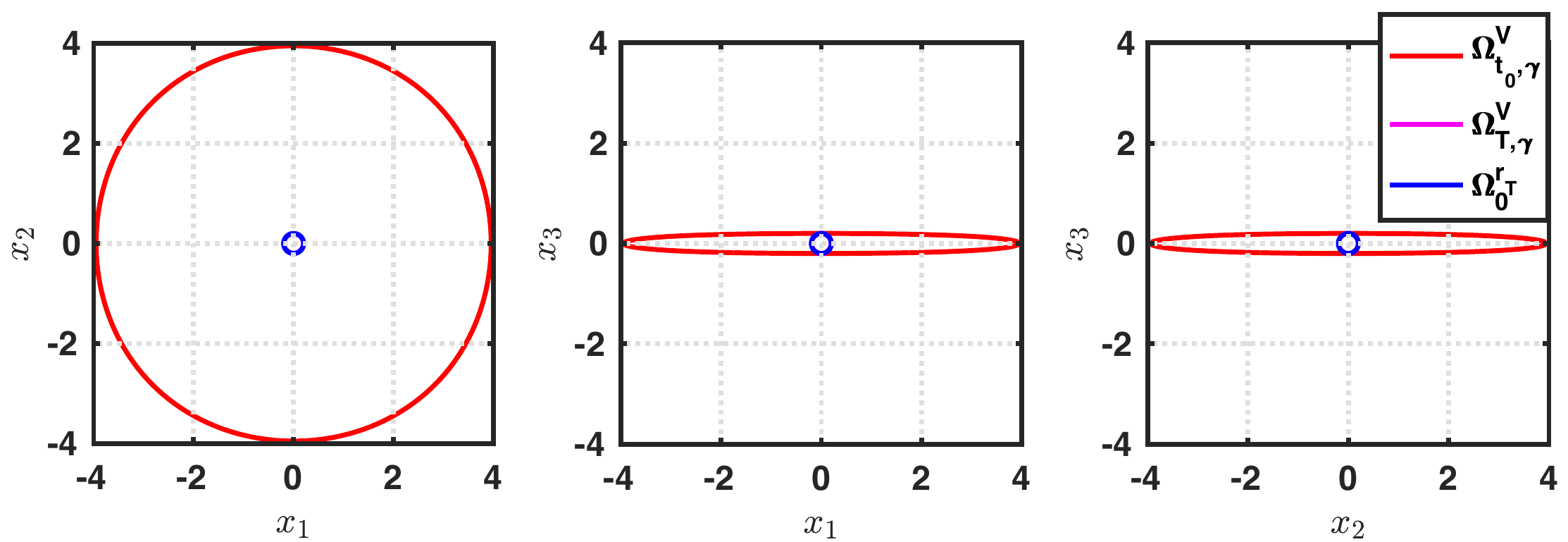}
	\caption{Dubin's car example}
	\label{fig:car}    
\end{figure}

Consider Dubin's car \cite{Dubins:57}, a multi-input system
\begin{align}
\dot{a} = v \cos(\theta), \ \dot{b} = v \sin(\theta), \ \dot{\theta} = \omega, \nonumber
\end{align}
with states $a$: $x$ position, $b$: $y$ position, $\theta$: yaw angle and control inputs $\omega$: turning rate, $v$: forward speed. By a change of coordinates, it can be transformed into polynomial dynamics \cite{David:07}
\begin{align}
\dot{x}_1 = u_1, \ \dot{x}_2 = u_2, \ \dot{x}_3 = x_2 u_1 - x_1 u_2, \nonumber 
\end{align}
with $x_1 = \theta$, $x_2 = a \cos(\theta)+b \sin(\theta)$, $x_3 = -2(a \sin(\theta) - b \cos(\theta))+\theta x_2$, and $u_1 = \omega$, $u_2 = v - \omega*(a \sin(\theta) - b \cos(\theta))$. Assume time horizon $[t_0, T] = [0, 1 \ \text{sec}]$, and target set $\Omega_0^{r_T} = \{x \in \mathbb{R}^3 | x^T diag(1/0.2^2, 1/0.2^2, 1/0.2^2) x - 1 \leq 0\}$. Figure \ref{fig:car} show the slices of sets with $x_3 = 0$, $x_2 = 0$, and $x_1 = 0$, respectively.
%\begin{figure}[h]
%	\centering
%	\begin{subfigure}[b]{0.2\textwidth}
%		\centering
%		\includegraphics[width=1.2\textwidth]{figures/carx1x2.eps}
%		\caption{$x_1 - x_2$ plane}  
%		\label{fig:carA}
%	\end{subfigure}
%	\hfill
%	\begin{subfigure}[b]{0.2\textwidth}  
%		\centering 
%		\includegraphics[width=1.2\textwidth]{figures/carx1x3.eps}
%		\caption{$x_1 - x_3$ plane}
%		\label{fig:carB}
%	\end{subfigure}
%	\begin{subfigure}[b]{0.3\textwidth}  
%		\centering 
%		\includegraphics[width=0.8\textwidth]{figures/carx2x3.eps}
%		\caption{$x_2 - x_3$ plane}
%		\label{fig:carC}
%	\end{subfigure}
%	\caption{Dubin's car example }
%	\label{fig:car}    
%\end{figure}

\subsection{Cart-pole Example}\label{ex3}
The polynomial dynamics for cart-pole is from \cite{Julian:12}
\begin{align}
\bmat{\dot{x}_1 \\ \dot{x}_2 \\ \dot{x}_3 \\ \dot{x}_4} = \bmat{x_3 \\ x_4 \\ f_3(x_2, x_4) \\ f_4(x_2,x_4)} + \bmat{0 \\ 0 \\ g_3(x_2) \\ g_4(x_2)}u \nonumber
\end{align}
with 
\begin{align}
f_3(x_2,x_4) &= 0.11707 x_2^5 + 0.03591 x_2^3 x_4^2 - 1.6032 x_2^3 - 0.17201 x_2 x_4^2 + 3.0313 x_2, \nonumber \\
f_4(x_2,x_4) &= 0.24902 x_2^5 + 0.13049 x_2^3 x_4^2 - 5.6188 x_2^3 - 0.29147 x_2 x_4^2 + 23.9892 x_2, \nonumber \\
g_3(x_2) &= 0.02905 x_2^4 - 0.11289 x_2^2 + 0.3955, \nonumber \\
g_4(x_2) &= 0.096371 x_2^4 - 0.54277 x_2^2 + 0.7831, \nonumber 
\end{align}
where $x_1$ to $x_4$ represent $d$: distance of the cart from the origin, $\theta$: angle of the pole from the vertical position, $v$: speed of the cart, $\dot{\theta}$: angular velocity of the pole, respectively. Control input $u$ is the horizontal force applied to the cart. Polynomial dynamics are obtained by approximating the system using least squares for $x_2 \times x_4 \in [-\frac{\pi}{2}, \frac{\pi}{2}] \times [-\frac{3\pi}{2}, \frac{3\pi}{2}]$. 

\begin{figure}[h]
	\centering
	\includegraphics[width=0.7\textwidth]{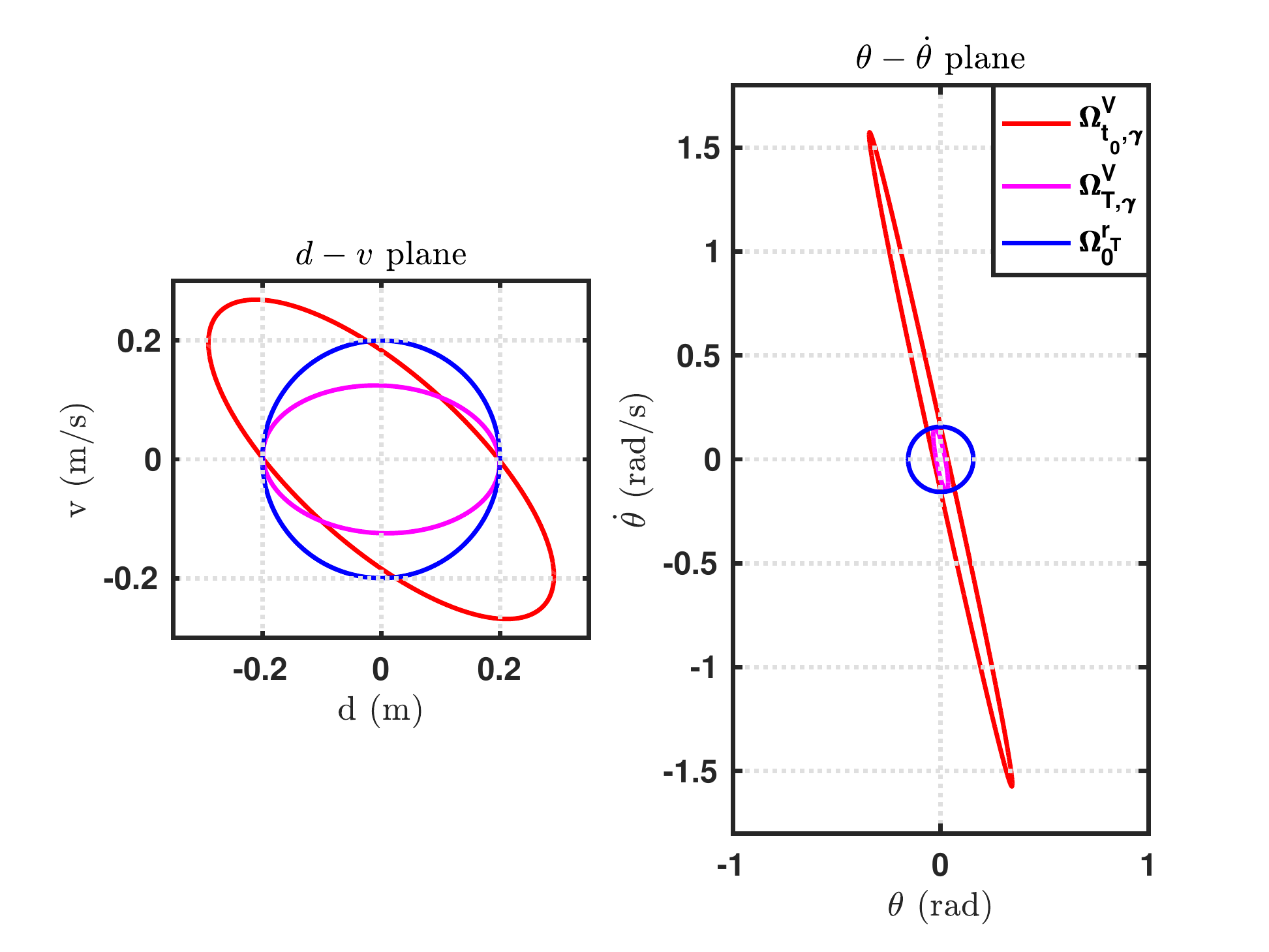}
	\caption{Cartpole example}
	\label{fig:cartpole}    
\end{figure}

Time horizon is $[0, 1 \ \text{sec}]$ and target set is $\Omega_0^{r_T} = \{x \in \mathbb{R}^4 |$ $x^T diag(1/0.2^2,$ $1/(\pi/20)^2,$ $1/0.2^2,$ $1/(\pi/20)^2) x - 1 \leq 0\}$. The sets shown on the left side of Figure \ref{fig:cartpole} are plotted with $\theta$ and $\dot{\theta}$ set to $0$. The sets shown on the right side of Figure \ref{fig:cartpole} are plotted with $d$ and $v$ set to $0$. 

\begin{figure}[h]
	\centering
	\includegraphics[width=0.5\textwidth]{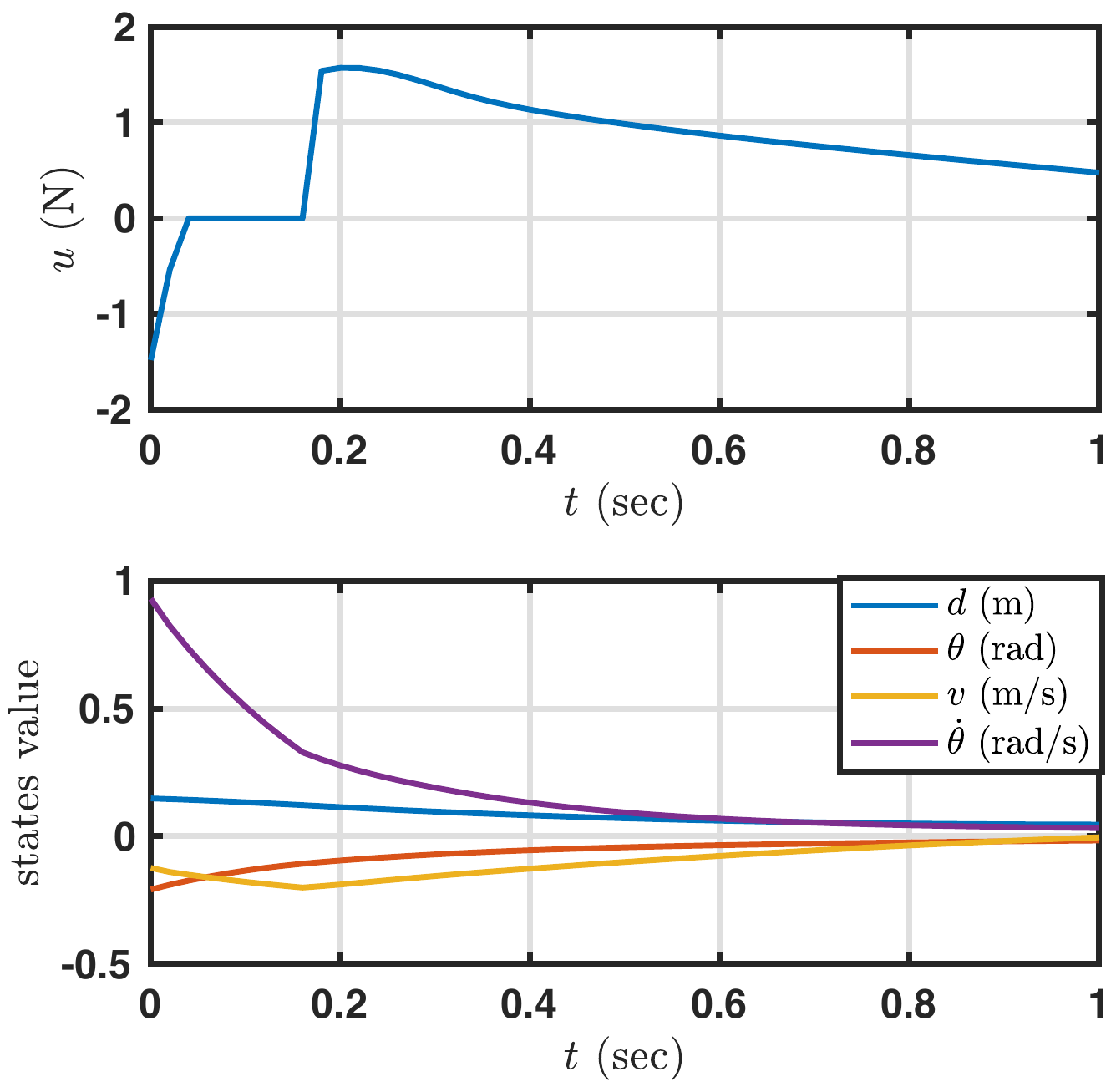}
	\caption{Cartpole simulations}
	\label{fig:cartpoleSim}    
\end{figure}

One simulation result is shown in Figure \ref{fig:cartpoleSim}, with initial condition $[0.148 \ \text{m},$ $-0.2088 \ \text{rad},$ $-0.1242 \ \text{m/s},$ $0.9301 \ \text{rad/s}]^T$, under the controller given by equations (\ref{eq:ab1})(\ref{eq:controlLaw}). 

\subsection{Pendubot Example}\label{ex4}
Consider the following polynomial dynamics for the pendubot
\begin{align}
\bmat{\dot{x}_1 \\ \dot{x}_2 \\ \dot{x}_3 \\ \dot{x}_4} = \bmat{x_2 \\ f_2(x_1,x_2,x_3,x_4) \\ x_4 \\ f_4(x_1,x_2,x_3,x_4)} + \bmat{0 \\ g_2(x_3) \\ 0 \\ g_4(x_3)}u \nonumber
\end{align}
with 
\begin{align}
f_2 &= - 10.6560 x_1^3 + 11.5309 x_1^2 x_3 + 7.8850 x_1 x_3^2 + 0.7972 x_2^2 x_3  + 0.8408 x_2 x_3 x_4 + 21.0492 x_3^3  + \nonumber \\
&\ \ \ \ \ 0.4204 x_3 x_4^2 + 66.5225 x_1 - 24.5110 x_3, \nonumber \\
f_4 &= 10.9955 x_1^3 - 48.9151 x_1^2 x_3 - 6.4044 x_1 x_3^2 - 2.3955 x_2^2 x_3  - 1.5943 x_2 x_3 x_4 - 51.9088 x_3^3 - \nonumber \\
&\ \ \ \ \ 0.7971 x_3 x_4^2 - 68.6419 x_1 + 103.9783 x_3, \nonumber \\
g_2 &= -10.0959 x_3^2 + 44.2521, \nonumber \\
g_4 &=  37.8015 x_3^2 - 83.9120, \nonumber
\end{align}
which is obtained as a least-squares approximation of the full equations for $x_1 \times x_3 \in [-1, 1] \times [-1, 1]$.

Here $x_1$ and $x_3$ represent $\theta_1$ and $\theta_2$, which are angular positions of the first link and the second link (relative
to the first link), respectively, and $x_2$ and $x_4$ are $\dot{\theta}_1$ and $\dot{\theta}_2$, which are angular velocities of the first and second link respectively. Input $u$ is the torque applied at the joint of first link and ground. 

\begin{figure}[h]
	\centering
	\includegraphics[width=0.7\textwidth]{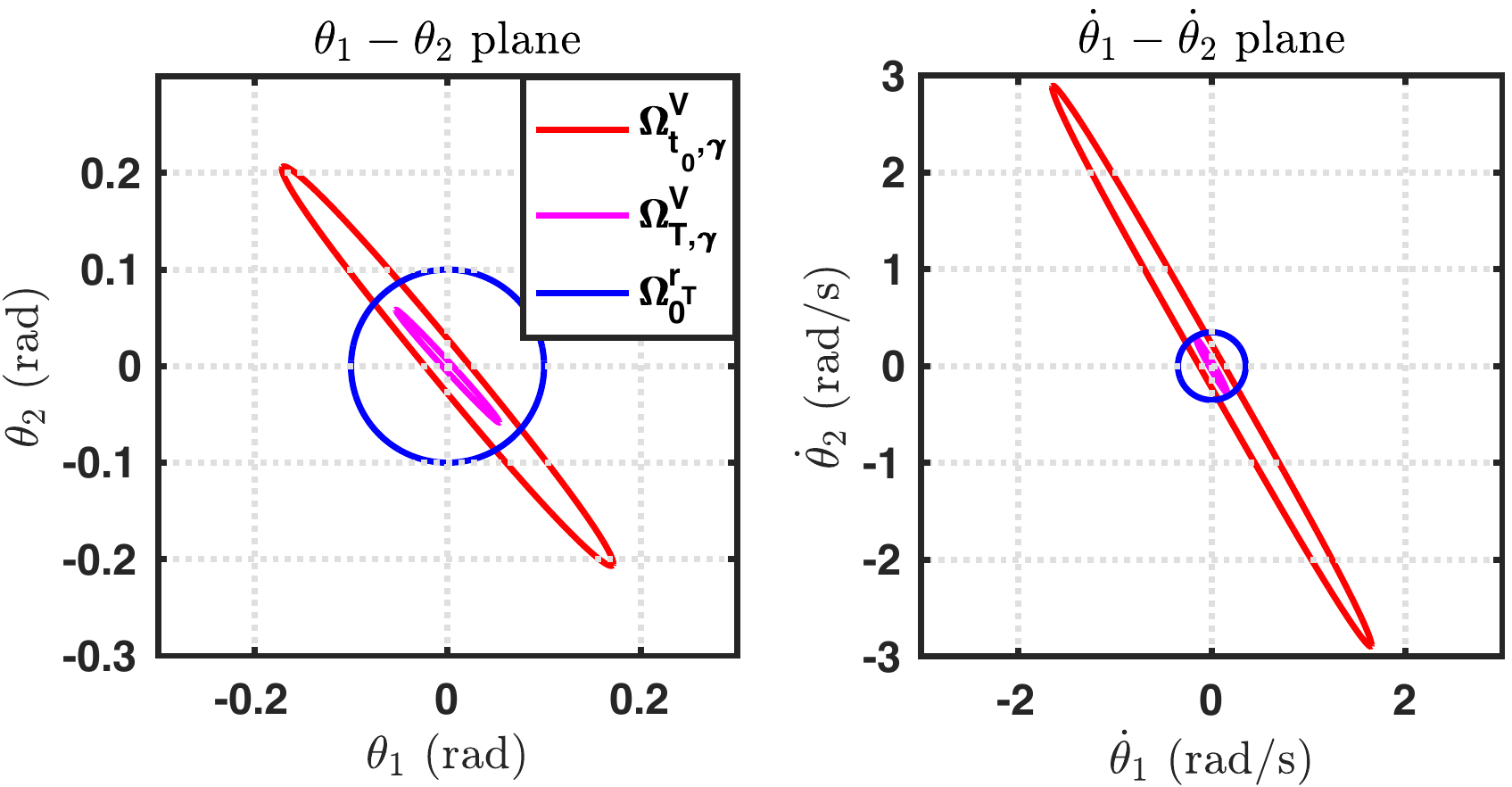}
	\caption{pendubot example}
	\label{fig:pendubot}    
\end{figure}

Time horizon is $[0, 1 \ \text{sec}]$ and $\Omega_0^{r_T} = \{x\in \mathbb{R}^4 | x^T$ $diag(1/0.1^2,$ $1/0.35^2,$ $1/0.1^2,$ $1/0.35^2) x - 1$ $\leq 0\}$. Sets shown on the left side of Figure \ref{fig:pendubot} are plotted with $\dot{\theta}_1$ and $\dot{\theta}_2$ set to $0$. Sets shown on the right side of Figure \ref{fig:pendubot} are plotted with $\theta_1$ and $\theta_2$ set to $0$. 

\subsubsection{Pendubot with $\mathcal{L}_2$ Disturbance}
Assume that the pendubot system is disturbed by a $\mathcal{L}_2$ disturbance $w$ satisfying $\norm{w}_{2,T} \leq R = 0.015$ rad. In addition, we have apriori knowledge that $\int_{0}^t w^T(\tau) w(\tau) d\tau \leq R^2 q(t) = R^2t^2/T^2$, and $\norm{w(t)}_2 \leq 0.0212$ rad, for all $t \in [0, 1 \ \text{sec}]$,
\begin{align}
\bmat{\dot{x}_1 \\ \dot{x}_2 \\ \dot{x}_3 \\ \dot{x}_4} = \bmat{x_2 \\ f_2(x_1,x_2,x_3,x_4) \\ x_4 \\ f_4(x_1,x_2,x_3,x_4)} + \bmat{0 \\ g_2(x_3) \\ 0 \\ g_4(x_3)}u + \bmat{0\\0\\0\\1}w. \nonumber
\end{align}

\begin{figure}[h]
	\centering
	\includegraphics[width=0.5\textwidth]{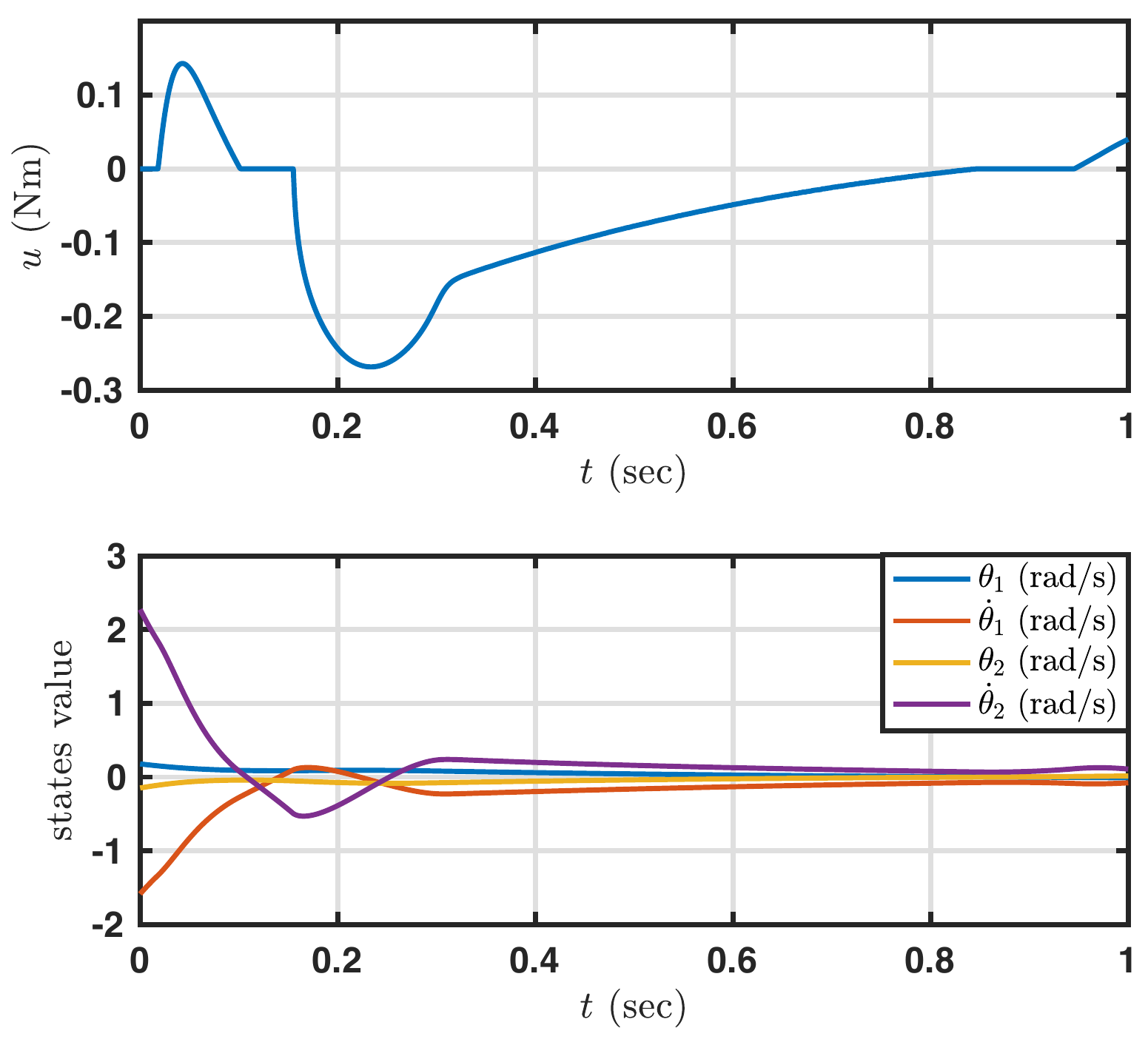}
	\caption{Disturbed Pendubot simulations}
	\label{fig:pendubotDisturbSim}    
\end{figure}

The simulation of pendubot system with control law from equations (\ref{eq:controlLaw})(\ref{eq:ab3}) and a disturbance signal $w(t) = \frac{\sqrt{2t}R}{T} \eta(t)$ is shown in Figure \ref{fig:pendubotDisturbSim}, where $\eta(t)$ is the value drawn from the uniform distribution on the interval $(0, 1)$ at each time step.

\subsection{NASA's Generic Transport Model (GTM) around straight and level flight condition}\label{ex5}
The GTM is a remote-controlled 5.5\% scale commercial aircraft. The open-loop longitudinal dynamics of the GTM \cite{Chakraborty:2011} is approximated as a degree-3, 4-state, 1-input polynomial system, where states are $U$ (m/s): air speed, $\alpha:$ angle of attack, $q:$ pitch rate, $\theta:$ pitch angle, and the input is $\delta_{elev}:$ elevator deflection (all angles expressed in radians).

\begin{figure}[h]
	\centering
	\includegraphics[width=0.7\textwidth]{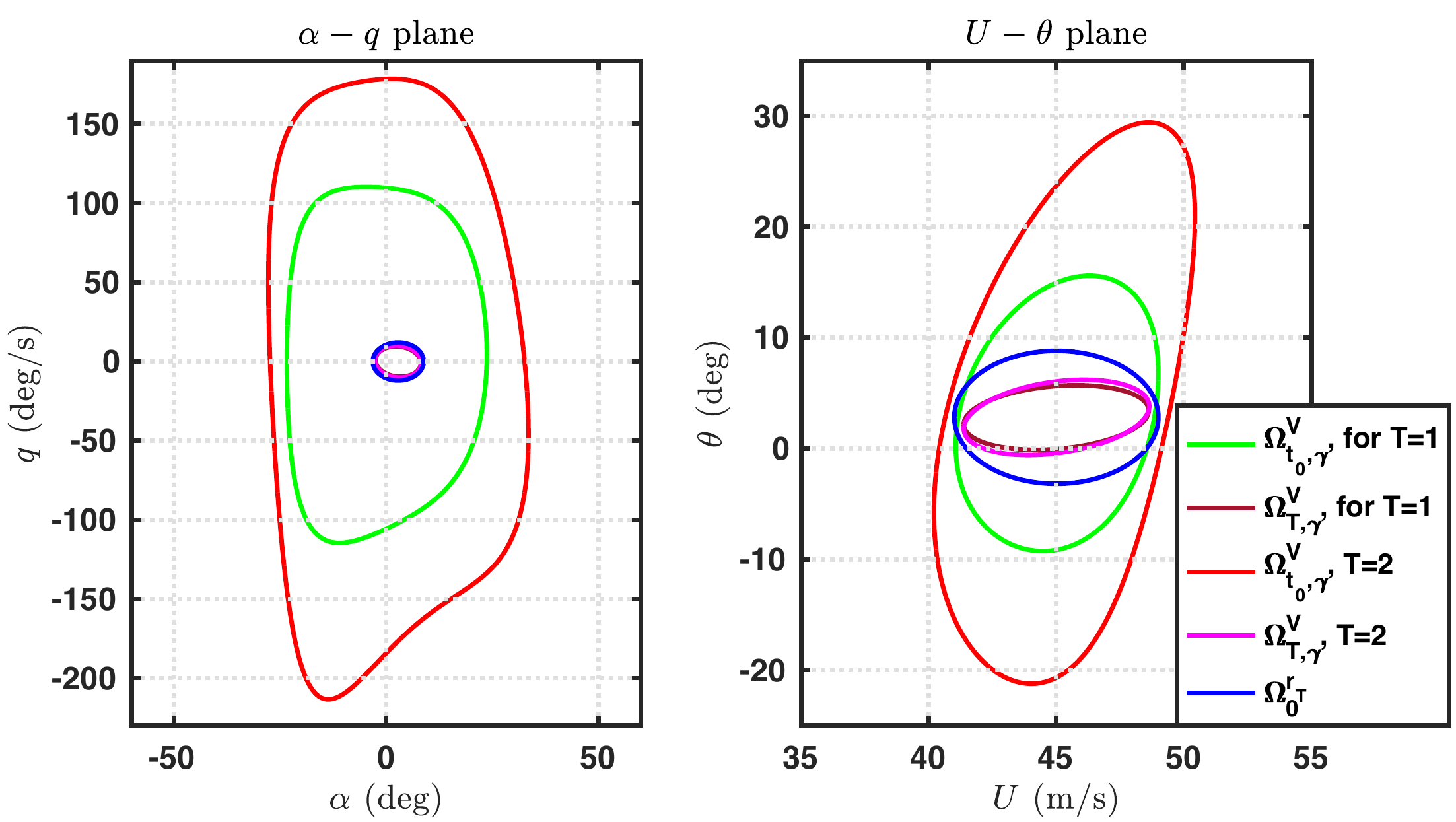}
	\caption{GTM example}
	\label{fig:GTM}    
\end{figure}

Given two time horizons [$0, 1$sec], [$0, 2$sec], and the target set $\Omega_{0}^{r_T} = \{x \in \mathbb{R}^4 | (x-x_{eq})^T diag(1/4^2, 1/(\pi/30)^2, 1/(\pi/15)^2, 1/(\pi/30)^2) (x - x_{eq}) \leq 1 \}$, where the equilibrium point $x_{eq} = [45, 0.04924, 0, 0.04924]^T$ represents the flight condition at level flight. The result shown in Figure \ref{fig:GTM} are slices of set at equilibrium point.

%Figure \ref{fig:GTMgamma} shows when using Algorithm \ref{alg1} to solve for storage functions, how $\gamma$ grows along with the iterations. We can see that after around 25 iterations, the algorithm converges.
%\begin{figure}[h]
%	\centering
%	\includegraphics[width=0.4\textwidth]{figures/gamma.eps}
%	\caption{$\gamma$ - iterations}
%	\label{fig:GTMgamma}    
%\end{figure}

\section{Conclusions}
We proposed a method for synthesizing controllers for nonlinear systems with polynomial vector fields. The synthesis process yields a storage function that characterizes the under-approximated BRS and a control law that steers the trajectories to the given target set from the under-approximation of BRS on a finite horizon. An iterative algorithm is proposed to construct the storage function, which is derived based on SOS programming, and min-norm optimization is used to compute control policies. The synthesis framework is also extended to uncertain systems with bounded uncertainties and $\mathcal{L}_2$ disturbances. This method is applied to several practical robotics and aircraft models.

\section*{Acknowledgements}
This work was funded in part by the ONR grant N00014-18-1-2209.
%\begin{thebibliography}{99}
%\end{thebibliography}
\bibliography{reference} 
\bibliographystyle{plain}
\end{document}